\pdfoutput=1
\documentclass[conference]{IEEEtran}
\usepackage{amsmath}
\usepackage{graphicx}
\usepackage{cite}
\usepackage{amsmath,amssymb,amsfonts,amsthm}
\usepackage{bm}
\usepackage{textcomp}
\usepackage{xcolor}
\usepackage{xspace}
\usepackage{url}
\usepackage{stfloats}
\usepackage{booktabs}
\usepackage[T1]{fontenc}
\usepackage[utf8]{inputenc}
\newcommand{\BfPara}[1]{{\noindent\bf#1.}\xspace}
\newcommand{\bra}[1]{\ensuremath{\left\langle#1\right|}}
\newcommand{\ket}[1]{\ensuremath{\left|#1\right\rangle}}
\newcommand{\xt}{\tilde{\bm{x}}}
\begin{document}

\title{A Tutorial on Quantum Amplitude Embedding\\with Data Re-Uploading}

\author{
\IEEEauthorblockN{
\parbox[t]{0.45\textwidth}{
\centering
Hyojun Ahn\\
\textit{Department of Electrical \& Computer Engineering}\\
\textit{Korea University}\\
Seoul, Republic of Korea\\
hyojun@korea.ac.kr
}
}
\and
\IEEEauthorblockN{
\parbox[t]{0.45\textwidth}{
\centering
Joongheon Kim\\
\textit{Department of Electrical \& Computer Engineering}\\
\textit{Korea University}\\
Seoul, Republic of Korea\\
joongheon@korea.ac.kr
}
}
}

\maketitle

\begin{abstract}
Amplitude embedding lets a variational quantum model read many classical features with few qubits, but the model output is then only a quadratic form in the normalized input. Data re-uploading, which encodes the input repeatedly between trainable blocks, can be used to enrich such a model. Because every additional upload of an amplitude embedding repeats a costly state preparation, the gain that an upload buys has to be known before the gates are spent. This tutorial thus addresses the problem of explaining what data re-uploading adds to an amplitude-embedded model and how large the gain is. We express sequential re-uploading and parallel re-uploading with two copies in one notation and show that each upload raises the reachable polynomial order by two. Under one trainable circuit family and one budget of trainable parameters, simulation results show that re-uploading raises the test coefficient of determination from 0.689 to between 0.883 and 0.975 on targets of order four and from 0.337 to between 0.642 and 0.783 on targets of order six, on all ten seeds. On a quadratic target and on handwritten digits, one upload remains the best choice, and every additional upload lowers the gate error that the classifier tolerates. These results help designers match the number of uploads to the polynomial order of the task at hand.
\end{abstract}

\begin{IEEEkeywords}
Quantum machine learning, variational quantum circuit, amplitude embedding, data re-uploading, state preparation, quantum data encoding
\end{IEEEkeywords}

\section{Introduction}\label{sec:intro}
Learning-based models now control aerial vehicles and schedule satellites~\cite{ahn2026ton,ahn2026boostrl}.
Variational quantum models are parameterized circuits trained as functions of classical inputs~\cite{mitarai2018qcl,cerezo2021vqa,oh2026trajectories}.
Such models are candidate learners for these systems~\cite{park2024trends,kim2025leo,kim2026sagin}.
They are also studied for visual recognition and distributed learning~\cite{kwak2023qddl,roh2025fastqcnn,kim2025quantumeyes,park2025dqfl}.
In all of these settings the classical input first has to be written into a quantum state, and this encoding step decides which functions of the input the model can express~\cite{schuld2021effect}.
Amplitude embedding stores the features in the amplitudes of a quantum state, so the number of qubits grows only logarithmically with the number of features.
This makes it a common encoding pattern for high-dimensional inputs~\cite{schuld2020circuit,roh2025qaa}.
The price is a rigid function class, because the output of an amplitude-embedded model is only a quadratic form in the normalized input vector, that is, a weighted sum of products of two features~\cite{schuld2019feature}.

Data re-uploading encodes the input several times between trainable blocks, each encoding being one upload, and it is the standard remedy for a similar rigidity in angle-encoded models~\cite{perezsalinas2020reuploading}.
The theory of re-uploading is written for angle encodings~\cite{schuld2021effect,jerbi2023beyond}.
Hands-on tutorials on re-uploading also use angle encodings~\cite{penatapia2023didactic}.
For an amplitude embedding, each additional upload repeats a state preparation whose gate count is exponential in the qubit count~\cite{mottonen2005transformation,shende2006synthesis}.
Studies of generalization also warn that encoding layers are not free~\cite{caro2021encoding}.
Empirical comparisons of encodings change the preprocessing, the qubit count, or the learning model together with the encoding~\cite{hur2022qcnn,rath2024encoding}.
A recent empirical study applies re-uploading to amplitude-encoded classifiers on two small tabular datasets and reports gains that depend on the dataset~\cite{tudisco2026encoding}.
Which targets make re-uploading pay off for an amplitude-embedded model, and by how much, thus remains unexplained.

\begin{figure*}[!t]
\centering
\includegraphics{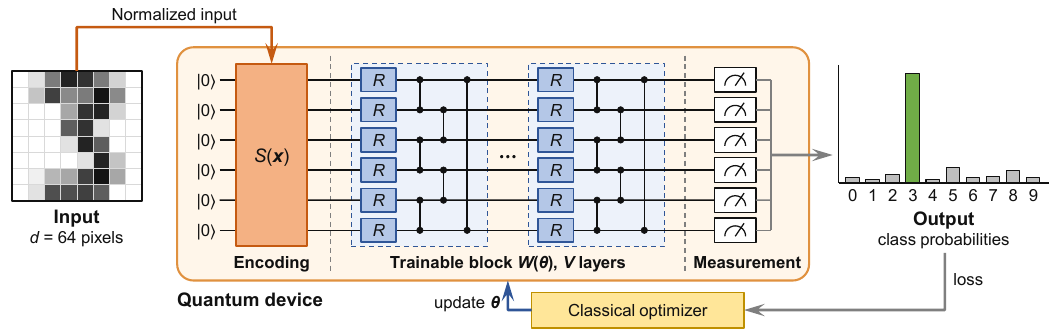}
\vspace{-2mm}
\caption{Amplitude-embedded variational quantum model with one upload. $S(\bm{x})$ writes the normalized image into six qubits, and an optimizer trains $\bm{\theta}$.}
\label{fig:model}
\vspace{-3mm}
\end{figure*}

We treat the state preparation of the amplitude embedding as the unit that is uploaded again.
A sequential scheme re-applies this state preparation between trainable blocks on the same qubits, and a parallel scheme prepares several copies of the embedded state side by side.
Viewed in the angles of the state preparation, every upload adds frequencies to the model, which raises by two the polynomial order of the input that the model can reach.
To attribute a gain to the uploads alone, all schemes share one trainable circuit family and one budget of trainable parameters, so that they differ only in the position at which the data enters the circuit.

The main contribution of this paper is a tutorial on data re-uploading for amplitude embedding that explains its fundamentals and quantifies its performance gain.
\begin{itemize}
\item We express amplitude embedding with sequential and with parallel re-uploading in one notation. We show that a sequential upload acts as an angle encoding in the state-preparation angles, so that each additional upload raises the reachable polynomial order by two and costs one more state preparation.
\item We measure the gain under one trainable circuit family, one budget of trainable parameters, and one training procedure, on synthetic targets of order 2, 4, and 6.
\item We delimit the gain. On a quadratic target and on handwritten digits, one upload remains the best choice, and every additional upload lowers the two-qubit gate error that the classifier tolerates.
\end{itemize}

\vspace{-1mm}
\section{Data Encoding in Variational Quantum Models}\label{sec:background}
This section reviews how classical data enters a variational quantum model, which Fig.~\ref{fig:model} illustrates for an image.
A register of $n$ qubits has $2^n$ basis states, and its state is a unit vector of $2^n$ amplitudes.
The model proceeds as follows.
\begin{enumerate}
\item The encoding normalizes the input and writes it into the amplitudes of the qubits with a state preparation $S(\bm{x})$.
\item The trainable block $W(\bm{\theta})$ applies $V$ layers, each of which rotates every qubit and entangles the qubits with a ring of controlled-Z gates.
\item The measurement returns outcome probabilities, from which the model output is computed.
\item A classical optimizer updates the trainable parameters from the loss, and steps 1) to 3) are repeated.
\end{enumerate}

A variational quantum model on $n$ qubits applies a circuit $U(\bm{x},\bm{\theta})$ that depends on the input $\bm{x}\in\mathbb{R}^d$ and on trainable parameters $\bm{\theta}$.
The model output is the expectation of an observable $O$, which can be expressed as,
\begin{equation}
f_{\bm{\theta}}(\bm{x})=\bra{0}U^{\dagger}(\bm{x},\bm{\theta})\,O\,U(\bm{x},\bm{\theta})\ket{0},
\label{eq:model}
\end{equation}
where $\ket{0}$ denotes the all-zero state of the $n$ qubits.
A classical optimizer trains $\bm{\theta}$ on a loss that is computed from $f_{\bm{\theta}}$~\cite{mitarai2018qcl}.
The circuit structure can also be searched, and such models have been combined with language models~\cite{son2026qnas,roh2026qdl}.

Amplitude embedding writes $d=2^n$ features into the amplitudes of $n$ qubits, and the encoded state is defined as,
\begin{equation}
\ket{\psi(\bm{x})}=\sum_{i=0}^{d-1}\tilde{x}_i\ket{i},\qquad \xt=\bm{x}/\lVert\bm{x}\rVert_2,
\label{eq:ae}
\end{equation}
where $\ket{i}$ denotes the $i$-th computational basis state and $\tilde{x}_i$ denotes the $i$-th entry of $\xt$.
A state-preparation unitary $S(\bm{x})$ maps $\ket{0}$ to $\ket{\psi(\bm{x})}$, and a trainable circuit $W(\bm{\theta})$ follows it, so that $U=W(\bm{\theta})S(\bm{x})$.
For a real input, the model output in \eqref{eq:model} can then be expressed as,
\begin{equation}
f_{\bm{\theta}}(\bm{x})=\xt^{\top}M_{\bm{\theta}}\,\xt,\qquad M_{\bm{\theta}}=\operatorname{Re}\!\left[W^{\dagger}(\bm{\theta})\,O\,W(\bm{\theta})\right],
\label{eq:quad}
\end{equation}
which is a quadratic form in the normalized input.
With two features, for example, the output can contain $\tilde{x}_0^2$, $\tilde{x}_0\tilde{x}_1$, and $\tilde{x}_1^2$ but no term of higher order.
Training changes the matrix $M_{\bm{\theta}}$ but not the degree, so a deeper $W$ adds no nonlinearity in $\bm{x}$.

\begin{figure}[!t]
\centering
\includegraphics[width=\columnwidth]{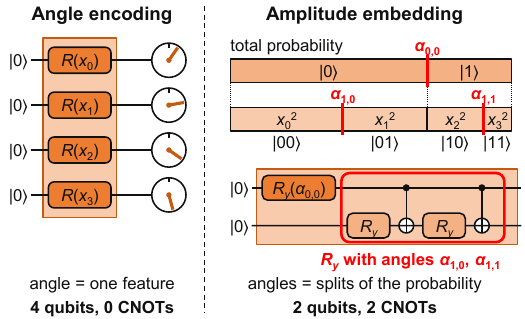}
\vspace{-2mm}
\caption{One upload of four features. Both encodings use rotations whose angles depend on the data. Angle encoding turns one qubit per feature, and each angle $\alpha_{k,c}(\bm{x})$ of amplitude embedding splits the probability mass.}
\label{fig:prep}
\vspace{-3mm}
\end{figure}
Angle encoding writes each feature as the angle of a single-qubit rotation, and Fig.~\ref{fig:prep} compares one upload of four features in the two encodings.
A model of this kind is a partial Fourier series in the data, that is, a sum of terms $\cos(mx)$ and $\sin(mx)$ with integer frequencies $m$, and the encoding gates determine which frequencies appear~\cite{schuld2021effect,oh2026fourier}.
Repeating the encoding gates between trainable blocks, which is called data re-uploading, enlarges the frequency set with every repetition.
The repetitions are not free, because generalization bounds grow with the number of encoding gates and more expressive circuits have smaller training gradients~\cite{caro2021encoding,mcclean2018barren}.

For angle encoding, one upload is a layer of single-qubit gates.
For amplitude embedding, one upload is a state preparation.
An $R_y(\alpha)$ gate rotates one qubit by the angle $\alpha$, and a CNOT gate flips one qubit when another qubit is one.
The standard construction is a cascade of uniformly controlled $R_y$ rotations, in which the rotation angle on qubit $k$ is selected by the $k$ preceding qubits~\cite{mottonen2005transformation}.
The rotation on qubit $k$ has one angle $\alpha_{k,c}$ for every setting $c=0,\dots,2^k-1$ of these qubits, and the angle is defined as,
\begin{equation}
\alpha_{k,c}(\bm{x})=2\arctan\sqrt{p_{k+1,2c+1}/p_{k+1,2c}},
\label{eq:angle}
\end{equation}
where $p_{k,c}$ denotes the sum of $\tilde{x}_i^2$ over block $c$ when $\xt$ is cut into $2^k$ blocks of consecutive entries, counted from zero.
The first rotation splits the total probability, which is the sum of the squared amplitudes, between the two halves of $\xt$, and each later rotation splits one of these parts again, until the parts in Fig.~\ref{fig:prep} are the squared amplitudes.
Each amplitude is therefore a product of $n$ factors $\cos(\alpha_{k,c}/2)$ or $\sin(\alpha_{k,c}/2)$, one for every split.
For example, $\tilde{x}_2=\sin(\alpha_{0,0}/2)\cos(\alpha_{1,1}/2)$.
A uniformly controlled rotation with $k$ control qubits takes $2^k$ CNOT gates, as Fig.~\ref{fig:prep} shows for one control~\cite{bergholm2005uniformly}.
Preparing a real nonnegative vector thus takes $2^n-2$ CNOT gates, and a general complex vector takes about twice as many~\cite{shende2006synthesis}.
This count is linear in the number of features $d$ and exponential in the number of qubits, a known caveat of loading classical data~\cite{aaronson2015fineprint}.
In both encodings an upload is thus a set of rotations whose angles depend on the data, and the encodings differ in these angles and in the gate count.

\vspace{-1mm}
\section{Re-Uploading an Amplitude Embedding}\label{sec:analysis}
\begin{figure*}[!t]
\centering
\begin{minipage}[b]{120mm}
\centering
\includegraphics{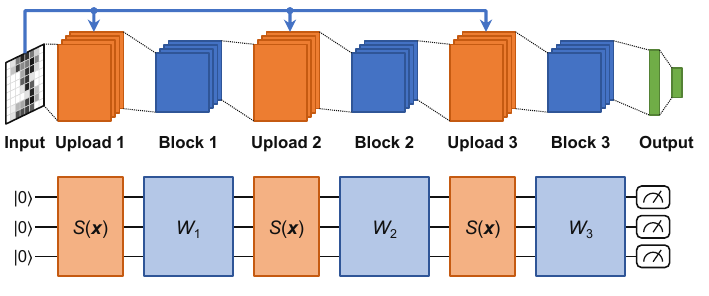}\\[1mm]
{\footnotesize (a) Sequential, $L=3$}
\end{minipage}\hfill
\begin{minipage}[b]{52mm}
\centering
\includegraphics{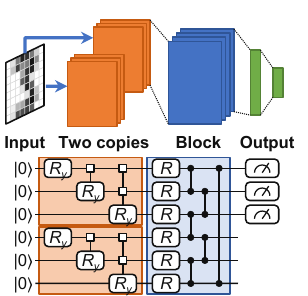}\\[1mm]
{\footnotesize (b) Parallel, two copies}
\end{minipage}
\caption{Re-uploading schemes for amplitude embedding. Orange blocks upload the input, and blue blocks are trainable.}
\label{fig:schemes}
\vspace{-3mm}
\end{figure*}
This section derives what a second upload adds to \eqref{eq:quad} and what it costs, as Table~\ref{tab:schemes} summarizes.
\begin{table}[!t]
\caption{Encoding schemes on $d=2^n$ features}
\label{tab:schemes}
\centering
\renewcommand{\arraystretch}{1.0}
\setlength{\tabcolsep}{3.5pt}
\begin{tabular}{@{}llll@{}}
\toprule
Scheme & Qubits & Enc.\ CNOTs & Model output \\
\midrule
Amplitude embedding & $n$ & $2^n-2$ & Quadratic form in $\xt$ \\
Sequential, $L$ uploads & $n$ & $L(2^n-2)$ & Frequencies $\le L$ in $\alpha_{k,c}$ \\
Parallel, $k$ copies & $kn$ & $k(2^n-2)$ & Degree-$2k$ polynomial in $\xt$ \\
Angle, $L$ uploads & $d$ & $0$ & Frequencies $\le L$ per feature \\
\bottomrule
\end{tabular}
\vspace{-3mm}
\end{table}
Fig.~\ref{fig:schemes} shows the two ways of repeating the upload of Fig.~\ref{fig:model}, each drawn as a block view above the corresponding circuit.

\BfPara{Sequential re-uploading}
The sequential scheme in Fig.~\ref{fig:schemes}(a) re-applies $S(\bm{x})$ between $L$ trainable blocks on the same $n$ qubits, and the circuit can be expressed as,
\begin{equation}
U_L(\bm{x},\bm{\theta})=W_L(\bm{\theta}_L)\,S(\bm{x})\cdots W_2(\bm{\theta}_2)\,S(\bm{x})\,W_1(\bm{\theta}_1)\,S(\bm{x}),
\label{eq:seq}
\end{equation}
where $W_l(\bm{\theta}_l)$ denotes the $l$-th trainable block.
Only the first upload prepares $\ket{\psi(\bm{x})}$.
The later uploads act on a state that is no longer $\ket{0}$, so they are data-dependent unitaries with no simple expression in $\xt$.
The cascade that implements $S(\bm{x})$ depends on the input only through the $R_y$ angles $\alpha_{k,c}(\bm{x})$ in \eqref{eq:angle}.
The circuit in \eqref{eq:seq} is therefore an angle-encoded model in these $2^n-1$ angles, and the Fourier analysis of re-uploading applies to it~\cite{schuld2021effect}.
One upload produces frequencies of at most one in each angle, which is the quadratic form in \eqref{eq:quad}.
With $L$ uploads the frequencies reach $L$.
A polynomial of degree $2L$ in $\xt$ generally contains the frequency $L$ in these angles.
A target of order $2L$ in the input thus needs at least $L$ uploads, whatever the depth of the trainable blocks.
The case $n=1$ makes this concrete.
Here $S(\bm{x})=R_y(\alpha)$ with $\xt=(\cos(\alpha/2),\sin(\alpha/2))$, so one upload gives $f_{\bm{\theta}}=a_0+a_1\cos\alpha+b_1\sin\alpha$ with coefficients that depend on $\bm{\theta}$.
A second upload adds $\cos 2\alpha$ and $\sin 2\alpha$, and $\cos 2\alpha=1-8\tilde{x}_0^2\tilde{x}_1^2$ is a quartic polynomial in $\xt$.

\BfPara{Parallel re-uploading}
The parallel scheme in Fig.~\ref{fig:schemes}(b) prepares $k$ copies of $\ket{\psi(\bm{x})}$ on $kn$ qubits and applies one trainable circuit $W(\bm{\theta})$ to all of them.
The copies are prepared from the same classical input, so no quantum state has to be cloned.
For two copies, the model output can be expressed as,
\begin{equation}
f_{\bm{\theta}}(\bm{x})=\bra{\psi(\bm{x})}\!\bra{\psi(\bm{x})}\,W^{\dagger}(\bm{\theta})\,O\,W(\bm{\theta})\,\ket{\psi(\bm{x})}\!\ket{\psi(\bm{x})},
\label{eq:par}
\end{equation}
which is a quadratic form in $\xt\otimes\xt$ and hence a homogeneous polynomial of degree four in $\xt$.
With $k$ copies the degree is $2k$.
This is the circuit counterpart of the polynomial kernel that copies of an embedded state provide~\cite{schuld2019feature}.
Two copies thus reach the same order as two sequential uploads at a different cost in qubits and depth.

\BfPara{Cost}
Table~\ref{tab:schemes} sets the function class of each scheme against its resources, namely the number of qubits (Qubits) and the CNOT gates spent on encoding nonnegative inputs (Enc.\ CNOTs).
The sequential scheme keeps $n$ qubits and multiplies the encoding count by $L$.
The preparations run in series, so the circuit depth grows by one state preparation per upload.
The parallel scheme multiplies the count by $k$ as well but spreads the gates over $k$ registers that are prepared side by side, at the price of $kn$ qubits.
With $n=6$, one upload costs 62 CNOT gates, whereas one trainable layer of the circuits in Section~\ref{sec:eval} costs only six two-qubit gates.
These counts decide whether a quantum learner fits platforms on which learned controllers operate, such as aerial and underwater vehicles~\cite{ahn2025hagpt,ahn2026access,kim2026aircraft}.

\vspace{-1mm}
\section{Performance Evaluation}\label{sec:eval}
The evaluation measures how much re-uploading improves an amplitude-embedded model and where the improvement ends.
Section~\ref{sec:analysis} predicts a gain only for targets of higher order than two, so the first task controls the order of the target.
The order of a target is its polynomial degree in the input.
The second task is an image benchmark, and a final analysis prices the additional state preparations under gate errors.

\vspace{-1mm}
\subsection{Experimental Setup}\label{sec:setup}
\begin{table}[!t]
\caption{Experimental settings of the two tasks}
\label{tab:setup}
\centering
\renewcommand{\arraystretch}{1.0}
\setlength{\tabcolsep}{4pt}
\begin{tabular}{@{}lcc@{}}
\toprule
Setting & Synthetic targets & Digits \\
\midrule
Features $d$ / qubits $n$ & 8 / 3 & 64 / 6 \\
Trainable layers $V$ & 36 & 24 \\
Trainable parameters $3nV$ & 324 & 432 \\
Training / test samples & 4,000 / 2,000 & 100 to 1,297 / 500 \\
Training steps & 2,400 & 600 \\
Learning rates searched & \multicolumn{2}{c}{0.003, 0.01, 0.03, 0.1} \\
Reported / pilot seeds & \multicolumn{2}{c}{10 / 2} \\
Metric & Test $R^2$ & Test accuracy \\
\bottomrule
\end{tabular}
\vspace{-3mm}
\end{table}
\BfPara{Models}
Table~\ref{tab:setup} lists the settings.
All models use one trainable circuit family, that is, one ansatz.
A layer applies a general single-qubit rotation with three parameters to every qubit and then a ring of controlled-Z gates, as in Fig.~\ref{fig:model}.
We compare amplitude embedding with one upload ($L=1$), sequential re-uploading with $L\in\{2,3,4\}$, and parallel re-uploading with two copies.
The total number of trainable layers $V$ is fixed and is split evenly over the $L$ blocks, and the two-copy model uses $V/2$ layers on $2n$ qubits.
Every model therefore has the same $3nV$ trainable parameters, and the models differ only in where the data enters.
The state preparation $S(\bm{x})$ is the cascade of uniformly controlled $R_y$ rotations, which is exact for nonnegative inputs.

\begin{figure}[!t]
\centering
\includegraphics[width=\columnwidth]{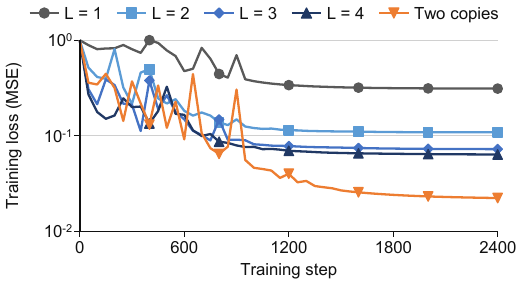}
\vspace{-4mm}
\caption{Training loss on the order-4 target (ten-seed mean).}
\label{fig:conv}
\vspace{-3mm}
\end{figure}
\begin{figure}[!t]
\centering
\includegraphics[width=\columnwidth]{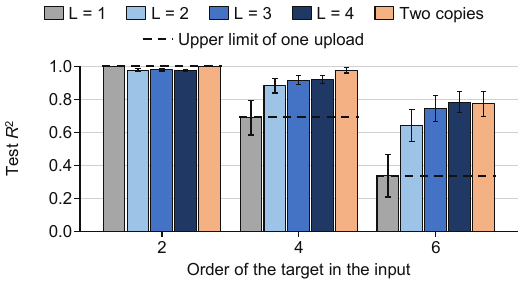}
\vspace{-4mm}
\caption{Test $R^2$ on synthetic targets of order 2, 4, and 6 (mean and standard deviation over ten seeds). The dashed line is the best quadratic form.}
\label{fig:order}
\vspace{-3mm}
\end{figure}

\BfPara{Tasks}
The synthetic task isolates the order of the target.
Inputs are uniformly distributed on the nonnegative part of the unit sphere.
The target is a product of one, two, or three standardized random quadratic forms in $\xt$, which gives the orders 2, 4, and 6 that Fig.~\ref{fig:order} compares.
We set $V=36$, so that every trainable block of the three-qubit models has at least 81 parameters, more than the 63 of a general three-qubit unitary, and the trainable part is not the bottleneck.
The model output is a trained linear combination of the eight outcome probabilities.
The metric is the test coefficient of determination $R^2$, which is one for a perfect fit and zero for predicting the mean.
The image task uses the $8\times 8$ handwritten digits distributed with scikit-learn, with 1,797 images in ten classes~\cite{pedregosa2011sklearn}.
The 64 pixel intensities are embedded in six qubits, and $V=24$ leaves six layers per block with four uploads.
Four qubits are measured, and the predicted class is the most probable of their first ten outcomes~\cite{baek2023logdim}.
The training loss is their cross-entropy scaled by one trained factor.

\BfPara{Training}
All models are trained with the Adam optimizer.
Training is full-batch with a cosine learning-rate schedule.
Runs take 2,400 steps on the synthetic task, by which point the training loss in Fig.~\ref{fig:conv} has leveled off.
On the digits, 600 steps suffice for every model to fit its training set.
Initial parameters are uniform in $[-\pi,\pi]$.
The learning rate of each model is selected on held-out training data of two pilot seeds, which are not among the ten reported seeds.
A seed fixes the data split or target and the initial parameters, and all models see the same data for a given seed.
We report means and standard deviations over the ten seeds, together with the number of seeds on which each model beats $L=1$.

\vspace{-1mm}
\subsection{Convergence}\label{sec:conv}
Fig.~\ref{fig:conv} shows that the gain of re-uploading is already present in the training loss and is not an effect of the stopping point.
The loss is the mean squared error on the standardized target.
One upload levels off at 0.312, whereas two, three, and four uploads level off at 0.109, 0.073, and 0.064, and two copies reach 0.022.
The models with $L=1$ to $L=4$ are within 5\% of their final loss after at most 1,400 of the 2,400 steps, so more steps would not let one upload catch up.

\vspace{-1mm}
\subsection{Gain on Targets of Higher Order}\label{sec:order}
Fig.~\ref{fig:order} shows that re-uploading lifts an amplitude-embedded model above a limit that one upload cannot pass.
The dashed line is the best quadratic form in $\xt$, obtained by least squares, and \eqref{eq:quad} makes it the upper limit of $L=1$.
On the order-4 target, one upload reaches an $R^2$ of 0.689 against a limit of 0.694, so further training of $L=1$ has nothing left to gain.
Two uploads raise $R^2$ to 0.883, three and four uploads reach 0.916 and 0.918, and two copies reach 0.975.
The paired gains over $L=1$ are 0.194, 0.227, 0.229, and 0.286, and each model is ahead on all ten seeds.
Measured against the gap that one upload leaves to a perfect fit, two uploads close 62\% of the gap and two copies close 92\%.
The two-copy model leads because its function class is the set of degree-4 polynomials in Table~\ref{tab:schemes}, which contains the target.

On the order-6 target, the gain follows the number of uploads as Section~\ref{sec:analysis} predicts.
One upload gives 0.337, which again sits at the limit of a quadratic form.
Two uploads give 0.642, three give 0.745, and four give 0.783, which are gains of 0.305, 0.408, and 0.446.
Every model is again ahead of $L=1$ on all ten seeds, and four uploads close 67\% of the gap that one upload leaves.
The gain thus keeps growing with every upload when the order of the target is high, whereas it saturates after three uploads on order 4.
Two copies give 0.771 and are no longer the best model, because their function class ends at degree four.
On the order-2 target, one upload already reaches 0.999, and re-uploading has nothing to add.
The sequential models are slightly lower on all ten seeds, between 0.975 and 0.977, and two copies reach 0.997.
The model classes are not nested, because the later uploads act on states other than $\ket{0}$.
A pure quadratic form is thus harder to realize once uploads interrupt the trainable layers.

\vspace{-1mm}
\subsection{Limits of the Gain}\label{sec:limits}
\begin{table}[!t]
\caption{Test accuracy (\%) on the digits (mean $\pm$ standard deviation)}
\label{tab:digits}
\centering
\renewcommand{\arraystretch}{1.0}
\setlength{\tabcolsep}{3.2pt}
\begin{tabular}{@{}lrccc@{}}
\toprule
Model & 2q gates & 100 images & 400 images & 1,297 images \\
\midrule
Amplitude, $L=1$ & 206 & \textbf{87.4 $\pm$ 1.9} & \textbf{94.0 $\pm$ 1.1} & \textbf{95.9 $\pm$ 1.1} \\
Sequential, $L=2$ & 268 & \underline{\smash{84.2 $\pm$ 2.6}} & \underline{\smash{91.8 $\pm$ 0.9}} & \underline{\smash{94.8 $\pm$ 0.8}} \\
Sequential, $L=3$ & 330 & 81.7 $\pm$ 1.9 & 89.4 $\pm$ 1.4 & 92.3 $\pm$ 1.2 \\
Sequential, $L=4$ & 392 & 68.9 $\pm$ 2.4 & 84.2 $\pm$ 2.0 & 88.5 $\pm$ 1.4 \\
Parallel, two copies & 268 & 82.2 $\pm$ 2.6 & 90.7 $\pm$ 2.8 & 94.3 $\pm$ 0.7 \\
\bottomrule
\end{tabular}
\vspace{-3mm}
\end{table}
\begin{figure}[!t]
\centering
\includegraphics[width=\columnwidth]{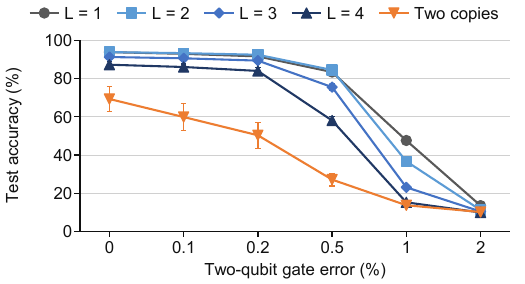}
\vspace{-4mm}
\caption{Digit accuracy with 1,024 shots against the two-qubit gate error (mean and standard deviation over ten seeds).}
\label{fig:noise}
\vspace{-3mm}
\end{figure}
Table~\ref{tab:digits} shows that the gain does not carry over to a task that one upload already fits, where bold and underline mark the best and the second-best model.
With all 1,297 images, one upload reaches 95.9\%, two uploads reach 94.8\%, three reach 92.3\%, and four reach 88.5\%.
The deficit grows as the training set shrinks, and four uploads lose 7.4 percentage points with 1,297 images and 18.5 points with 100 images.
Two copies lose less, 1.6 points with 1,297 images and 5.2 points with 100 images.
The cause is generalization and not the fit, since every model classifies at least 99.9\% of its training images correctly, which agrees with bounds that grow with the number of encoding gates~\cite{caro2021encoding}.
The digits thus behave like the order-2 target, on which additional uploads add capacity that the task does not need.
The ranking of the models does not depend on the selected learning rates.
With a common rate of 0.01, four uploads reach 75.1\% with 100 images and 88.8\% with 1,297 images and remain below one to three uploads.

The second limit is the hardware cost of the additional state preparations, which Fig.~\ref{fig:noise} measures.
Under a global depolarizing channel, we evaluate the classifiers that are trained on 1,297 images, and the circuit fidelity is calculated as,
\begin{equation}
F=(1-\epsilon)^{G},
\label{eq:fid}
\end{equation}
where $\epsilon$ denotes the two-qubit gate error rate and $G$ denotes the two-qubit gate count in Table~\ref{tab:digits}.
The outcome distribution is then the ideal one with weight $F$ and the uniform one with weight $1-F$.
A shot is one run of the circuit followed by a measurement.
We draw 1,024 shots per image, repeat the sampling 20 times, and vary $\epsilon$ from 0 to 2\%.
This range covers the 0.62\% reported for two-qubit gates on a 53-qubit processor~\cite{arute2019supremacy}.
Training is noiseless, so the results measure how much error a trained circuit tolerates.
Fig.~\ref{fig:noise} shows that each upload lowers the gate error that the classifier tolerates, and that the two-copy model is limited by shot noise before gate errors.
Without gate errors, 1,024 shots lower $L=1$ from 95.9\% to 93.7\% and $L=2$ from 94.8\% to 93.8\%, so both models are level.
One upload loses more because its margin between the two most probable classes is 0.058 on average, against 0.085 for $L=2$.
At $\epsilon=1\%$ the 62 additional CNOT gates of the second upload dominate, and $L=1$ keeps 47.6\% against 36.7\% for $L=2$.
Four uploads fall off first, from 87.2\% without gate errors to 58.0\% at $\epsilon=0.5\%$, where their fidelity is 0.14 against 0.36 for $L=1$.
The two-copy model reaches only 69.2\% with 1,024 shots, far below its 94.3\% in exact evaluation.
Its average margin is 0.013, less than a quarter of that of $L=1$, so its classes are resolved only with many more shots.
The number of uploads should therefore follow the polynomial order of the task, and each upload beyond that order costs both accuracy and two-qubit gates.

\vspace{-1mm}
\section{Conclusion}\label{sec:conclusion}
In this paper, we have presented the fundamentals of data re-uploading for amplitude embedding and quantified its performance gain under one trainable circuit family and one budget of trainable parameters.
We have shown that a sequential upload acts as an angle encoding in the state-preparation angles.
A target of order $2L$ in the input therefore needs at least $L$ uploads, each at $2^n-2$ CNOT gates.
On synthetic targets of order 4 and 6, re-uploading has raised the test $R^2$ from 0.689 to as much as 0.975 and from 0.337 to as much as 0.783, on all ten seeds.
On a quadratic target and on handwritten digits, one upload has remained the best choice, and every additional upload has lowered the tolerated gate error.
These results come from simulated circuits of three to twelve qubits with nonnegative inputs and noiseless training.
They still give designers a rule, namely to match the number of uploads to the polynomial order of the task.

\vspace{-1mm}
\section*{Acknowledgements}
This research was supported by MSIT (Ministry of Science and ICT), Korea, under ITRC (Information Technology Research Center) support program (IITP-2026-RS-2024-00436887) supervised by IITP (Institute for Information \& Communications Technology Planning \& Evaluation); and this work was also supported by the National Research Foundation (NRF) by the Korea Ministry of Science and ICT (Information and Communications Technology) (MSIT) (RS-2025-00561377). \textit{(Corresponding author: Joongheon Kim)}

\bibliographystyle{IEEEtran}
\bibliography{refs}

\end{document}